\documentclass[sigconf]{acmart}

\AtBeginDocument{%
  \providecommand\BibTeX{{%
    \normalfont B\kern-0.5em{\scshape i\kern-0.25em b}\kern-0.8em\TeX}}}

\copyrightyear{2023}
\acmYear{2023}
\setcopyright{rightsretained}
\acmConference[IUI '23]{28th International Conference on Intelligent User Interfaces}{March 27--31, 2023}{Sydney, NSW, Australia}
\acmBooktitle{28th International Conference on Intelligent User Interfaces (IUI '23), March 27--31, 2023, Sydney, NSW, Australia}
\acmDOI{10.1145/3581641.3584064}
\acmISBN{979-8-4007-0106-1/23/03}

\usepackage{graphicx}  
\usepackage{url}
\usepackage{booktabs}  
\usepackage{caption}
\usepackage{subcaption}
\usepackage{tabularx}

\graphicspath{{figures/}{pictures/}{images/}{./}}

\begin{document}


\title[Addressing UX Challenges in Designing ML]{Addressing UX Practitioners’ Challenges in Designing ML Applications: an Interactive Machine Learning Approach}




\author{K. J. Kevin Feng}
\affiliation{%
  \institution{University of Washington}
  \city{Seattle}
  \country{USA}}
\email{kjfeng@uw.edu}

\author{David W. McDonald}
\affiliation{%
  \institution{University of Washington}
  \city{Seattle}
  \country{USA}}
\email{dwmc@uw.edu}

\renewcommand{\shortauthors}{Feng and McDonald}

\begin{abstract}
  UX practitioners face novel challenges when designing user interfaces for machine learning (ML)-enabled applications. Interactive ML paradigms, like AutoML and interactive machine teaching, lower the barrier for non-expert end users to create, understand, and use ML models, but their application to UX practice is largely unstudied. We conducted a task-based design study with 27 UX practitioners where we asked them to propose a proof-of-concept design for a new ML-enabled application. During the task, our participants were given opportunities to create, test, and modify ML models as part of their workflows. Through a qualitative analysis of our post-task interview, we found that direct, interactive experimentation with ML allowed UX practitioners to tie ML capabilities and underlying data to user goals, compose affordances to enhance end-user interactions with ML, and identify ML-related ethical risks and challenges. We discuss our findings in the context of previously established human-AI guidelines. We also identify some limitations of interactive ML in UX processes and propose research-informed machine teaching as a supplement to future design tools alongside interactive ML.
\end{abstract}

\ccsdesc[500]{Human-centered computing~Empirical studies in interaction design}

\keywords{Interactive machine learning, interactive machine teaching, contextual inquiry, UX practice}

\maketitle

\section{Introduction}
Machine learning (ML) advancements have triggered a shift in the current technological landscape towards increasing inclusion of ML in user-facing applications \cite{amershi2019guidelines}. Indeed, ML can be seen woven into the fabric of everyday life, facilitating traffic predictions as we commute, correcting spelling as we write and work, providing biometrics as we exercise, and recommending music and films as we relax.

User experience practitioners\footnote{The field of user experience covers a diverse set of roles in the product development process---interaction design, user research, and project management are a few examples. Commonly, individuals in these roles map identifiable user needs to product features and capabilities, but rarely specialize in the writing of production-grade code \cite{hartson2012ux}. We use ``practitioners'' to indicate individuals whose primary responsibilities are in these user experience roles.} (UXPs) work to conceptualize, design, and prototype how users experience new ML powered applications. Through an iterative process of identifying user needs, translating the needs to task flows and graphical interfaces, and evaluating the interfaces with users, UXPs align user needs with the capabilities of the technology \cite{moore2017conversational, subramonyam2021protoai}. ML-enabled interfaces, however, give rise to novel design challenges for UXPs: ML capabilities can be ambiguous and changing, ML outputs can be unpredictable, and ML systems can make errors \cite{yang2020difficult}.


ML has been described as a new design material with unique properties \cite{dove2017material, yang2018material, luciani2018material, benjamin2021material, subramonyam2021towards, holmquist2017intelligence}. Specifically, ML's technical abstractions (which are typically removed from use context), probabilistic nature, and capability uncertainty make it a difficult design material to work with \cite{dove2017material, subramonyam2021towards, benjamin2021material}. Additionally, the infrastructure to support designing with ML is still not well-developed. There is both a lack of ML education for designers \cite{dove2017material} and a lack of prototyping tools that enable designers to directly ``play around'' with ML \cite{yang2020difficult}. 
Little research has enabled designers to shape model specifications according to user needs in early-stage ML development.


Efforts to make ML more accessible to non-experts\footnote{Our definition of ``non-experts'' is based on Yang et al. \cite{yang2018imt}: those who are not formally trained in ML but still build ML solutions for practical tasks.} delivers promising potential in this area. HCI researchers have contributed numerous tools and techniques in the domain of interactive machine learning (IML) to allow non-experts to experiment directly with datasets and models (e.g. \cite{subramonyam2021towards, teachable-machine, facets, lobe, ramos2020imt, fogarty2008cueflik, amershi2012regroup, mishra2021transfer}). 
However, much of the work with non-experts targets end-users rather than UXPs. Because UXPs play a mediating role between the technology and end-user, it is not guaranteed that the tools resolving end-user challenges with ML will also resolve challenges faced by UXPs \cite{false-consensus}. In fact, we have yet to see concrete evidence of IML's benefits to UXPs in practice, despite prior work positing that those benefits exist \cite{dove2017material, yang2020difficult}. This brings us to our motivating question: \textit{in what ways does direct model experimentation via IML help address UXPs' challenges of working with ML as a design material?}

We used Google's Teachable Machine \cite{teachable-machine}---an interactive, no-code IML model-building interface for non-experts---to conduct a task-based design study with 27 UXPs to answer this question. 
Unlike previous literature, we provided UXPs the opportunity to experience a simplified ``end-to-end'' ML pipeline that allowed them to change the training task and model classes in addition to testing model outputs. We found that providing UXPs with hands-on experience with IML in their UX workflows mitigates many ML design challenges mentioned by previous work. Namely, UXPs were able to align ML models and underlying data to user goals, derive design affordances to enhance end-user interactions with ML, and foresee ethical risks and challenges that can come with embedding ML into their work. Just within the span of our design sessions, UXPs were able to develop recognition of key human-AI guidelines, even if they had no prior ML experience nor knowledge of the guidelines. Given this, IML-enabled design tools may be appealing at first glance. Yet, we observed that many UXPs' mental models and goals for ML were more well-aligned with the paradigm of interactive machine teaching (IMT) \cite{ramos2020imt}. We establish a new model of IMT tailored specifically for UXPs, which we dub ``\textit{research-informed machine teaching} (RIMT)'' to address the incongruity and shortcomings of IMT when used in the context of UX practice. We discuss how IML and RIMT can exist harmoniously in future design tools. To summarize, our main contributions in this work are:

\begin{enumerate}
    \item A collection of insights from UXPs reflecting sophisticated understanding of ML as a design material upon hands-on experience with IML. 
    \item Analytical takeaways on intuitive recognition of human-AI guidelines via combining UX expertise with IML.
    \item Research-informed machine teaching, a conceptual guide for UXPs designing ML-enabled interfaces.
\end{enumerate}

\section{Related Work}

\subsection{ML as a Design Material}
\label{s:rr-ml-material}
Applying a \textit{material framework} to unfamiliar technologies can be appealing for designers \cite{subramonyam2021towards}. From an industrial design perspective, Doordan \cite{doordan2003materials} proposes a material framework for design consisting of 1) \textit{fabrication}---preparation of materials with specific properties for initial use, 2) \textit{application}---transforming materials into usable products, and 3) \textit{appreciation}---gathering responses to the material from user communities. While this framework was created with physical materials in mind, Robles and Wiberg argue that the advances in computational materials bring about a ``material turn'' that also instills materiality as a vital consideration in digital materials \cite{robles2010material}. 

Prior work has shown that ML is difficult to grasp as a design material \cite{dove2017material, yang2018material, subramonyam2021towards, benjamin2021material, holmquist2017intelligence}. To illustrate why, we mapped challenges raised in prior work to Doordan's material framework\footnote{As an alternative, Yang et al. \cite{yang2020difficult} mapped UXPs' challenges with ML on the double diamond design process \cite{design-council-2005}.}:
\begin{itemize}
    \item \textit{Fabrication:} ML is typically prepared by ML practitioners (data scientists, ML engineers, etc.) and its properties are expressed as technical abstractions divorced from user-centric concepts designers are accustomed to \cite{subramonyam2021towards, subramonyam2022leaky, yang2018material}. This leaves designers lacking sufficient knowledge about the ML's capabilities and limits, often treating it as a black box \cite{yang2018experienced}.
    \item \textit{Application:} Because ML is viewed as a black box, it is difficult to define and calibrate user expectations to ML's (often unpredictable) behaviour \cite{yang2020difficult, benjamin2021material}. Designers are also unable to creatively manipulate the material, an essential activity when generating design solutions \cite{beaudouin2009prototyping, design-council-2005, giaccardi2015materials}. Further, designers are concerned about ethical and fairness issues ML may cast upon users \cite{dove2017material, holmquist2017intelligence}. 
    \item \textit{Appreciation:} ML may be constantly evolving in response to user inputs, but because of a lack of material understanding, it is challenging to reason about the nature of those evolutions \cite{yang2018material, yang2020difficult}. Additionally, some of ML's capabilities may only be uncovered through certain interactions or feedback, tangling aspects of \textit{fabrication} with \textit{appreciation} that then introduces friction within designers' workflows \cite{subramonyam2021towards}.
\end{itemize}

Researchers and practitioners have attempted to resolve design challenges with workflows and tools that combine model exploration with UI prototyping tasks \cite{subramonyam2021protoai}, probes that investigate non-expert understanding of ML evaluations \cite{oh2020evluations}, process models \cite{subramonyam2021towards} and abstractions \cite{subramonyam2022leaky} that facilitate collaboration between designers and ML experts, and educational materials for designers \cite{hebron2016, pair-guidebook, hax-guidelines}. 

We observe an unresolved bottleneck in the \textit{fabrication} stage that is a cause of many later challenges in the material framework---designers are removed from the fabrication process and cannot easily gain insight into ML's material properties. We specifically target this bottleneck in our work by providing designers and other UXPs hands-on experience fabricating prototypical ML models, and observing how the resulting experiential knowledge aids them with common challenges encountered when designing with ML.

\subsection{Democratizing ML for Developers and End-Users}
\label{s:rr-tools}
Lowering the barrier for developers and end-users 
to work with ML has been a significant focus in ML democratization efforts. In traditional ML workflows, a ML practitioner first defines (or receives) model requirements and then completes \textit{data-oriented} tasks (data collection, cleaning, and labeling), followed by \textit{model-oriented} tasks (feature engineering, model training, evaluation, deployment, and monitoring), with some feedback loops in between \cite{amershi2019software}. Approaches to automating aspects of this workflow---such as selecting model architectures \cite{laredo2020automatic, truong2019towards}, tuning hyperparameters \cite{fusi2018probabilistic}, and engineering features \cite{liu2020autofis, severyn2013automatic}---have been developed by the ML community under a paradigm known as automated machine learning (AutoML) \cite{karmaker2021automl}. Major cloud providers of AutoML \cite{ms-aml, google-aml, h20-aml, ibm-aml, databricks-aml, amazon-aml} believe that the paradigm can enable ML non-expert developers and semi-expert ``citizen data scientists'' to create fully-fledged, deployable models, often with little to no code \cite{databricks-aml, ms-aml, google-aml, ibm-aml}. However, studies of AutoML usage in practice have shown that the users of AutoML systems are still primarily expert data scientists \cite{xin2021whither, crisan2021automl} and that data scientists are concerned about the harms arising from non-expert use of such tools \cite{crisan2021automl}. Additionally, AutoML models still remain as blackboxes and can only be created when large pools of labeled data are available \cite{ramos2020imt}. 

Besides AutoML, another major paradigm that makes model-building more accessible is interactive machine learning (IML). Dudley and Kristensson \cite{dudley2018iml} characterize IML as ``an interaction paradigm in which a user or user group iteratively builds and refines a mathematical model [...] through iterative cycles of input and review.'' Amershi et al. \cite{amershi2014power} add that the cycles should be rapid, focused, and incremental, while Fails and Olsen \cite{fails2003iml} adopted the term ``human(s)-in-the-loop'' to highlight human input and guidance throughout the ML workflow. The IML subfield of \textit{active learning} has been of interest to the HCI community for the model's ability to dynamically query a human to collect feedback and label new datapoints \cite{settles2012active, chao2010transparent, monarch2021human, raghavan2006active, settles2011theories}, achieving higher accuracy with fewer labeled examples \cite{settles2012active, settles2011theories}. In a similar vein of minimizing data requirements, Mishra and Rzeszotarski \cite{mishra2021transfer} designed an interface for \textit{transfer learning} to allow non-expert users to transfer learned representations from a larger model to a separate, domain-specific task. Generally, advances in IML have resulted in user-friendly, no-code tools that allow those with no ML experience to train a model in just a few clicks, including Google's Teachable Machine \cite{teachable-machine}, Lobe \cite{lobe}, and Liner \cite{liner}. These tools have primarily been used in education, but also for accessibility and creative tinkering \cite{carney2020teachable}.

The aforementioned paradigms seek to algorithmically extract knowledge from data. However, Ramos et al. \cite{ramos2020imt} argue that in order for models to simultaneously be intuitive for non-experts to build, incorporate domain expertise, and debuggable, learnable representations should directly come from human knowledge rather than implicit, data-derived knowledge. They introduce a process known as \textit{interactive machine teaching} (IMT) that leverages humans' inherent teaching capabilities to explicitly ``teach'' representations to models. IMT consists of 3 steps: 1) \textit{planning}---identification of a teaching task and a curriculum (set of materials to help teach the model, typically in the form of data), 2) \textit{explaining}---showing the learning agent examples and explicitly identifying concepts the agent should learn, and 3) \textit{reviewing}---correcting erroneous predictions and updating teaching strategy and/or the curriculum \cite{ramos2020imt}. Researchers have explored IMT and other flavours of machine teaching through knowledge decomposition strategies for teaching \cite{ng2020imt, sanchez2021train}, uncertainty perception \cite{sanchez2022teaching}, sensitizing concepts to guide the design of approachable IML tools \cite{yang2018imt}, building intelligent tutoring systems \cite{weitekamp2020imt}, and extending teaching to a visual format for computer vision tasks \cite{zhou2022gesture, sultanum2020teaching}.

We note that while current IML tools and techniques have made great strides toward supporting end-users and developers in working with ML, little attention has been given to UXPs. How can IML tools help UXPs mitigate well-documented challenges in designing ML-enabled interfaces, if at all? What unique considerations should IML tools make for UXPs over other users? The lack of studies on IML as a design aid leaves these open questions ripe for investigation. We address them in our study.

\section{Method}

Our task-based methods were motivated by contextual inquiry due to its ability to provide rich information about users' work practices and processes \cite{beyer1999contextual}. We note that while Beyer and Holtzblatt intended contextual inquiry to be performed in users' natural environment \cite{beyer1999contextual}, many people now work remotely from home offices due to COVID-19. Our study consisted of virtual 1-on-1 design sessions through Zoom with 27 industry UX practitioners (UXPs). In these sessions, we captured how they used Teachable Machine \cite{teachable-machine} to design and propose a proof-of-concept (POC) app that integrates a classifier as part of its core functionality. 

A key difference in our study design compared to those of prior work at the intersection of UX practice and ML (e.g. \cite{yang2018experienced, subramonyam2021protoai, subramonyam2021towards}) is that UXPs are tasked with crafting ML models themselves, rather than being provided a fully-trained model. We chose this method upon drawing from \textit{constructivism} in the learning sciences. Based on Piaget's theory of cognitive development \cite{piaget1976piaget}, constructivism claims that human knowledge is constructed as a result of interactions ``between a person's mental model and their experiential perceptions'' \cite{sarkar2016constructivist}. Since its formalization, constructivism has been applied to technical domains to educate novice programmers \cite{lister2011concrete} and designing IML systems \cite{sarkar2016constructivist}. In addition to applications to learners, Taber \cite{taber2012constructivism} argues that constructivism should also be applied to teachers via constructivist pedagogy. That is, teachers draw connections to students' prior learning and experiences and use on-going assessment to adjust teaching approaches in the light of how learners' needs \cite{taber2012constructivism}. We posit that UXPs can benefit from a constructivist approach to grasping ML as a design material in Teachable Machine through both 1) hands-on experience with creating ML models \textit{as a learner}, and 2) modifying model specifications based on performance and user needs \textit{as a teacher}.

\subsection{Study Structure}

Each study session was 2 hours long. The first 90 minutes (the Activity portion) consisted of a design activity where participants were guided through a tutorial of the tool, briefed on the design prompt and supplementary resources, and given time to create their proposal: a presentation artifact (e.g. slide deck, document) to show their POC to theoretical project stakeholders. In the remaining 30 minutes (the Interview portion), participants were asked questions about their experiences using the tool and some of the design decisions they made throughout the process. We structured our Activity portion off of the industry-standard double diamond design process \cite{design-council-2005}, but designed an abridged version of the process to fit within the study's time constraints. This abridged version specifically honed in on the transition from idea formulation to solution exploration (regions around where the first and second diamonds meet in the original process). We provided participants with some resources they may otherwise need to dedicate non-trivial amounts of time to set up themselves so they can focus on crafting their POC proposals. 

All sessions were conducted over Zoom. Participants were rewarded a \$40 gratuity in the form of an Amazon gift card after completion of the session. We recorded all Interview portions of the sessions and collected all proposals generated by participants for analysis. We piloted our study with two UXPs who were well-known to the team and iterated based on feedback. For example, we originally did not provide sample user research insights to the participant but included it after realizing it was necessary to bridge a gap in a typical UX workflow. 

\subsection{Participants}
We recruited participants through UX- and HCI-related Slack channels and mailing lists associated with our institution, groups for UX professionals on Discord, LinkedIn, and Twitter (as well as publicizing with our personal LinkedIn and Twitter accounts), and personal connections. Participants were eligible if they 1) have at least one year of professional work experience in UX, and 2) were currently employed as a UX professional at the time of the study. We kept recruitment open as we conducted design sessions until we reached saturation.

Out of our 27 participants, 25 were based in the US, 1 was in Europe, and 1 was in Asia. 21 were UX designers and 6 were UX researchers. Most were early in their careers: 14 had 1 -- 2 years of work experience, 8 had 3 -- 5 years, 2 had 6 -- 10 years, and 3 had 11+ years. Most (15) worked for large organizations of 1000+, while some (5) were from medium organizations of 201 -- 1000 and the rest (7) were from small organizations of < 200. Participants came to UX from diverse backgrounds: visual/industrial design (14), computing (8), social \& behavioural sciences (7), management (4), natural sciences \& math (3), architecture (2), humanities (2), and informatics (1).

Most of our participants (20) did not have prior experience designing with AI. The remaining 7 had varying levels of prior AI design experience. However, slightly over half (15) reported they had previous exposure to AI through various avenues such as employer workshops, university courses, and online tutorials. 

\subsection{Study Protocol}
After explaining the study and gaining consent, participants downloaded a folder containing all files they would need for the activity. The files consisted of a PDF with instructions for the main design task, along with training and evaluation images for Teachable Machine. We manually collected the images from Wikimedia Commons \cite{wikicommons} for the tutorial task and randomly sampled from the EPFL Food-11 dataset \cite{food-dataset} (which we chose for its size and class diversity) for the main design task. 

\begin{figure*}[h]
    \centering
    \includegraphics[width=1\textwidth]{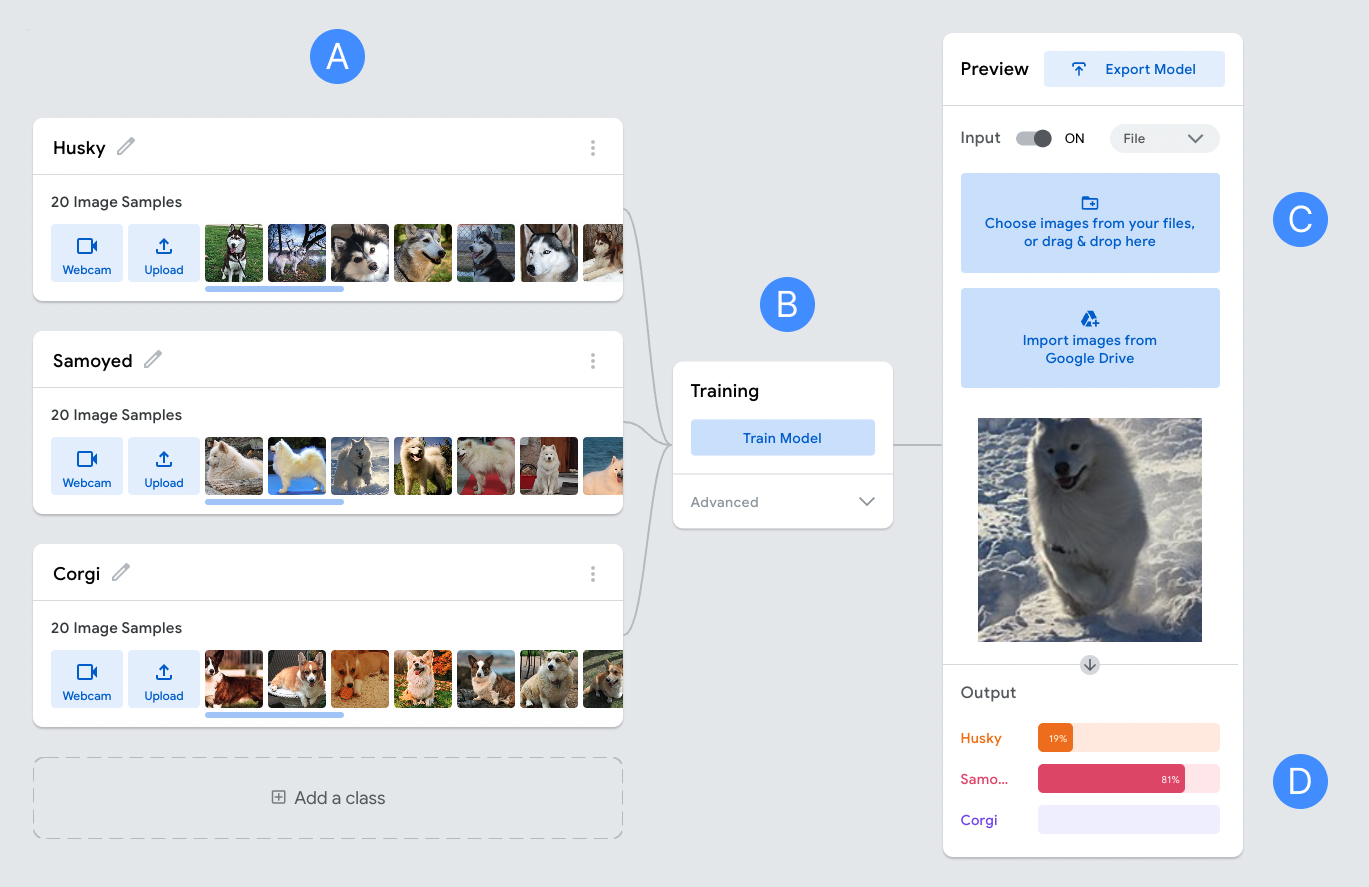}
    \caption{Overview of the Teachable Machine interface. \textbf{A}: class modules where users can drag-and-drop image files for upload. \textbf{B}: training module with a button for initiating model training, which typically takes less than 30 seconds. \textbf{C}: input module where users can upload an image for the model to evaluate. \textbf{D}: output module with the model's class probability scores on the evaluated image.}
    \label{f:tm-interface}
    \Description{The Teachable Machine interface against a gray background with white modules. 3 modules on the left represent classes (Husky, Samoyed, and Corgi) and contain 20 training images each. A training module in the middle features a blue  Train Model button. A module on the right allows users to upload evaluation images via Google Drive or local files. Here, an image of a samoyed running against a snowy backdrop is evaluated. The output features a bar chart visualization features detected probabilities for each class: 19\% for Husky, 81\% for Samoyed, and 0\% for Corgi.}
\end{figure*}

We selected Teachable Machine as the tool of choice for this study for 3 main reasons. First, it was free and publicly accessible in the browser and did not require participants to download any software or create new user accounts. Second, its visual layout and simplicity allowed non-technical users to quickly develop intuition for training ML models. The interface is shown in Fig. \ref{f:tm-interface}. While we recognize that the tool's simplicity restricted its ability to create highly sophisticated models, we accepted this trade-off as we prioritized usability for UXPs in our study. Third, and perhaps most importantly, the tool satisfied many themes for facilitating human-AI interaction design outlined by Yang et al. \cite{yang2020difficult}, as well as considerations for AI interface prototyping tools specified by Subramonyam et al. \cite{subramonyam2021protoai}. We experimented with other no-code ML platforms---including AutoML offerings from Google, Microsoft, and IBM---but found that the tools did not offer the same tight feedback loop for non-expert training, evaluation, and iteration of models that we considered to be essential for our study.

\subsubsection{Tutorial}
Participants were familiarized with Teachable Machine by creating two image classification models capable of distinguishing different breeds of dogs under a research team member's guidance. The training data consisted of 20 images of dogs belonging to each breed. We also provided test images consisting of two images from each dog breed. One of the test images was selected to trigger a misclassification---some participants noticed this while others did not. Participants first trained and evaluated a binary classification model, and then did the same for a 3-class model by adding another class on top of their previous model and retraining. All participants stated that they were comfortable with training an image model in Teachable Machine by the end of the tutorial. The length of the tutorial was typically around 10 minutes.

\subsubsection{Main Design Task}
Participants were presented with the following design prompt:
\begin{quote}\textit{Your company likes to invest in new ideas, particularly ones that use machine learning. You and another designer have an idea for a mobile app that uses machine learning to help users understand their eating habits. The basic idea is a photographic food journal to help users understand whether they are eating well. [...] You and your partner developed a preliminary persona to help you both stay focused on a potential user. You need to design and present a proof-of-concept for the app. }
\end{quote}

We created this prompt to 1) provide them with a concrete starting point, and 2) focus their attention on UX challenges associated with the ML aspects of the app. Since the main design activity was relatively short (slightly longer than an hour), we also fabricated user research insights and a persona for the participant to be able to quickly move to experimenting with models in Teachable Machine and creating their POC. We also acknowledge that there may be gendered biases around food and dietary habits. To mitigate this, we created both a female and male persona---keeping characteristics between them constant except for their photos, names, and background information. We counterbalanced the personas with 14 participants seeing the female persona and 13 seeing the male one. 

We provided participants with 3 datasets with which they can train and evaluate models in Teachable Machine. The datasets all contained the same 300 images randomly sampled from Food-11's 16,643 images, but were labelled in distinct ways: one had 2 classes, one had 3, and the other had 5. Each set of classes followed a specific mental model: 2-class was a representation of the healthy-unhealthy binary, 3-class was based on how restaurants categorize food on menus, and 5-class was in accordance with the MyPyramid food pyramid food groups \cite{britten2006development}, published by the USDA. We provided participants with the latest version of MyPyramid in case they were unfamiliar with it.

When sampling images from Food-11, we preserved the ratio of the number of images between classes to mimic the imbalance in the original dataset. One research team member then labelled and organized images from the original 11 classes into the new sets of 2, 3, and 5 target classes. The 3 datasets are described in Table \ref{t:datasets}. Participants were encouraged to train models using all 3 datasets, select the one they consider to be best-suited for the app, and use that for the rest of the activity.

\begin{table}[h]
\centering
\caption{Anatomy of the 3 datasets we provided to participants.}
    \begin{tabular}{p{3cm} p{2cm}}
    \toprule 
    Dataset & \# of images\\
    \midrule 
    
    \multicolumn{2}{l}{\textbf{2-class}} \\
    \quad healthy & 147 \\
    \quad unhealthy & 153 \\
    
    \multicolumn{2}{l}{\textbf{3-class}} \\
    \quad appetizers & 158 \\
    \quad entrees & 85 \\
    \quad desserts & 57 \\
    
    \multicolumn{2}{l}{\textbf{5-class}} \\
    \quad grains & 90 \\
    \quad vegetables & 47 \\
    \quad fruits & 14 \\
    \quad dairy & 42 \\
    \quad proteins & 107 \\
    
    \bottomrule
    \end{tabular}
    
    \label{t:datasets}
\end{table}

Finally, we briefed participants on guidelines for creating the proposal, which they would use to communicate the POC to stakeholders. We left the contents and authoring method of this proposal up to the participant; the only requirements we set were that it should be approximately 10 slides/pages, and that it needed to be exported as a PDF. 

\begin{figure*}[h]
    \centering
    \includegraphics[width=1\textwidth]{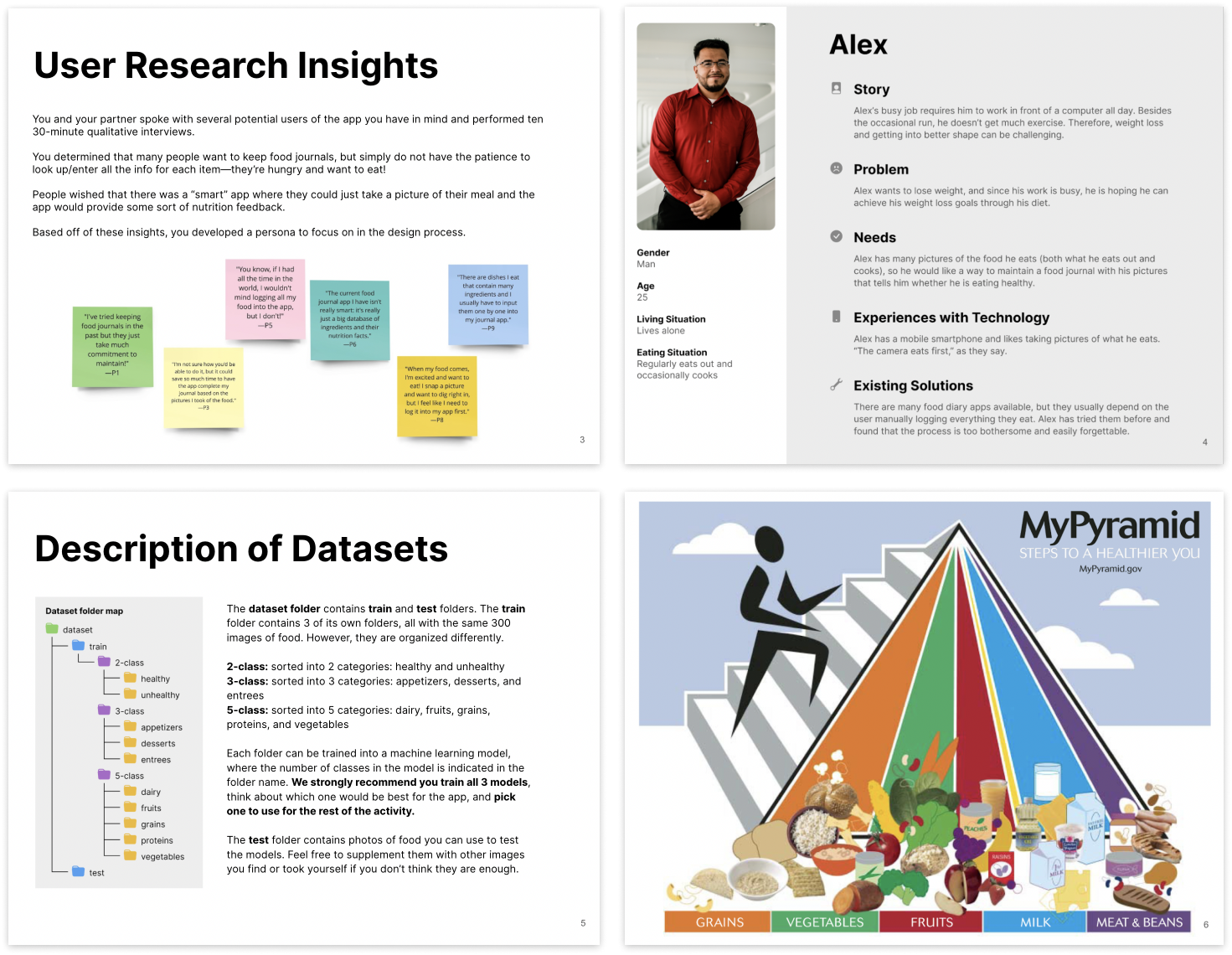}
    \caption{Sample pages from our PDF package we gave to participants. From left to right and top to bottom: user research insights, the male persona, datasets overview, MyPyramid food pyramid.}
    \label{f:pdf-pack}
    \Description{4 pages from the PDF package we provided to participants at the beginning of the study. The 4 pages feature user research insights we derived, a 25-year-old male persona named Alex, a description of datasets provided with a file tree diagram, and the MyPyramid food pyramid published by the USDA.}
\end{figure*}

All supplementary resources (design prompt, user research insights, persona, datasets guide, MyPyramid, and deliverable guidelines) were packaged into a PDF that the participant downloaded at the beginning of the session. Sample pages from the PDF can be seen in Fig. \ref{f:pdf-pack}, and the full document is available in our Supplementary Materials. After briefly walking participants through this PDF, we gave them time to work, checking in occasionally to answer questions and give notice of remaining time. This work period typically lasted 75 minutes. 

\subsubsection{Interview}
We conducted semi-structured interviews with our participants after they completed the main design activity. The discussion revolved around two main topics: Teachable Machine and the proposal. With regards to Teachable Machine, we asked them whether and how the tool enabled them to achieve the goals they envisioned for the POC, how the tool can better support them in understanding and communicating ML concepts they encountered, and general usability. We reviewed the proposals and asked participants about certain design decisions they made throughout the process (e.g. which dataset they chose to train their final model and why), as well as aspects of the process they thought were most important to communicate to stakeholders. We also asked participants about ethical challenges or risks they identified during the activity, how they may be mitigated, and next steps they may take to improve their work given more time and resources. Our full protocol is available in our Supplementary Materials. Our interviews typically lasted around 30 minutes.

\subsection{Data Analysis}
We focused our analysis around interview transcripts, deferring the analysis of the proposals to future work but still using them to contextualize interview dialogue. Interview transcripts were initially auto-generated from Zoom recordings, after which the first author manually reviewed the videos alongside the transcripts and corrected any mistranscribed areas. Two authors collectively took a first pass over the data, identifying insightful regions of the transcript that may be incorporated into further analysis. The first author then performed open and axial coding on those regions to identify themes and establish connections across themes. The themes were partially informed by frameworks in previous literature on AI user experience design and machine teaching (see Sections \ref{s:rr-ml-material} and \ref{s:rr-tools}).

\section{Results}


Through the task of developing models and proposing a POC, ML-enabled app, UXPs told us how AI can better serve end-users and tune user experiences accordingly with new AI-specific interactions. They also uncovered ethics and risks of using AI in their app and offered design suggestions for future IML tools tailored for UX practitioners. We elaborate on these areas below. Throughout this section, we switch to using ``AI'' over ``ML'' to better match participants' quotes, even though we acknowledge that all instances of AI were implemented via ML in our study.

\subsection{Reasoning About Data and AI Models}


\subsubsection{Some Presuppositions About AI Were Addressed by IML} 
\label{s:mental-models}
We initially asked practitioners about their expectations of AI prior to coming in for our study. A few thought of AI as a faraway concept, including ``futuristic sci-fi'' (P1) and ``one magical tool'' (P26). P7 specifically stated that thinking about how to incorporate AI into designs was ``something that's very overwhelming for me and I never really thought about it.'' In addition to their own assumptions, P11 also recognized that the public may see AI as ``dangerous, and somehow in the future control our lives.'' This influenced the techniques with which they designed their POC: ``I created an avatar for our AI [...] to make the AI seem [...] very friendly and close to your life and useful.'' 
Many recognized AI as imperfect, constantly evolving, and potentially harmful. Although P19 had no prior AI experience, they acknowledged recent popular dialogue on AI transparency and agreed that ``there's so much that we haven't necessarily [...] accomplished yet in terms of making [AI] safe for people to use and accurate.'' P19 extended this, mentioning the importance of ``educational measures in place to protect users against, or make them aware of, what the gaps are with [...] these technologies.'' From a UX standpoint, P15 stated concerns about over-reliance: ``I don't want to rely on it, and then set the user up for disappointment.'' Despite wariness, some participants expressed excitement about the UX improvements enabled by ML, and consequently wanted to learn some ML fundamentals: ``I see some good places to actually apply machine learning and improve the experience. So that's why I'm interested in this area, just trying to learn some basics [...] in my free time'' (P10).

Hands-on experience with IML helped participants calibrate their expectations about model capabilities and performance, validating some presuppositions about AI imperfections or bringing them to participants' attention if they were initially unaware. P1 summarizes the main issue addressed by IML:

\begin{quote}
\textit{``You know, in your head, like this thing can be classified in this way, but you don't have a good understanding of how often or in what ways it's going to error.''}
\end{quote}

P18 mentioned that the tight feedback loop was helpful in verifying hypotheses they had for the model: 

\begin{quote}
\textit{``When I was waiting for the result, it was pretty quick, like just within a few seconds, so I think that kind of like quick feedback is very helpful for me to know, well, am I doing this wrong or correctly, and then getting the result that I knew [to expect].'' }
\end{quote}

Indeed, the benefits of quickly testing and verifying hypotheses with IML were noted by many participants. P22 added that even though Teachable Machine and the provided datasets were too simple to be used in a production-ready app, it was still valuable to their designs to ``see what results would I get if I tried [the models].'' In addition to calibrating performance expectations, some participants also found that IML helped them understand and mitigate potential biases. Even going through the Teachable Machine tutorial, P1 said: ``You start to understand how bias could work like the example with the dogs and that one where it categorizes the black and white corgi as a husky [...] then you're already thinking about how you're training it and how you might remove bias.'' Later on in the POC, P1 suggested ``training with foods from as many cultures as possible, so that no one feels excluded'' could mitigate food-related biases.

In some cases, participants casted doubt on AI's ability to achieve what they want after initial assumptions were addressed. Some had a mental model of AI that would constantly receive feedback and improve dynamically over time, but were unable to experience that in the experimental tasks. Many wanted users to be shown more granular nutritional information, but were skeptical that AI can identify those details from images alone. As P6 stated, ``I'm not sure how efficient [the] model is in terms of identifying vitamins and minerals just from the food picture.'' Many pointed out that it would be very difficult for the model to recognize portion size, which can be essential in diet tracking. A couple participants (P22, P23) also pointed out that everyday nuances in food, such as sharing items with others or only eating a partial meal, can be challenging for AI to track personal food intake. A few also expressed general skepticism about AI's ability to deliver an acceptable experience for users, as excessive errors will easily ``lead to mistrust by users'' (P9). 
Just like many others, P14 realized that relying on AI alone was not enough and identified some important considerations before diving into any IML tool to start AI experimentation: 

\begin{quote}
    \textit{``The tool is one thing. I think the kind of the overall algorithm and how do we define it, what kind of outcome we want to achieve, I feel like that's the most important part, or that's something we should define before even before the tool.'' }
\end{quote}

That is, IML may \textit{augment} certain design and prototyping activities (e.g. better preparing UXPs to have conversations with AI experts), but it does not \textit{replace} any of them or anyone involved (e.g. the AI experts). 

\subsubsection{IML Enabled Exploration of App and User Goals}

Participants found it valuable to interactively explore different combinations of model classes we provided (see Table \ref{t:datasets}) to see which one should be used in the app. P6 stated: ``while doing that I kind of like at the back of my head thought, how would these classifications kind of benefit the user?'' Indeed, many found the exploration helpful in aligning classes (and consequently the model itself) with user needs. For example, P12 identified protein tracking as a desire for their persona and chose the 5-class model. P19 considered the binary classes of \textsc{healthy} and \textsc{unhealthy} to be overly interpretive: ``One thing that's deemed unhealthy for one culture or one country might be actually healthy for someone in another, so I [was] steering away from those kinds of interpretive labels.'' Many others agreed that the 5-class model was best for the app because it was the most flexible, most detailed, and least subjective. Despite many claiming that the 2-class model was overly judgemental and difficult to meet users' expectations of healthiness, a few participants considered it to be more ideal than the 5-class model due to its simplicity and straightforwardness. P7 thought it could better surface unhealthy eating habits: ``At the end of the day, I could always go back to my journal and if I wanted to figure out what my unhealthy habits, where I could [review] the pictures and maybe decipher something from that.'' P23 considers simplicity and ease of testing to be more important in a POC: ``We probably want to start with something minimal that we can actually test with users, and so I think it's easier to do that with the two category model.'' No participants thought the 3-class one should be used, as it seemed irrelevant and uninformative. P21 actually considered the 3-class model to be the most accurate, but avoided it due to lack of comprehensiveness: ``Where's breakfast, where's snacks? Also [the model] is extremely ethnocentric to a very small subset of the world.'' 

Exploring class combinations also gave participants new ideas for classes that might better align with user goals. A common suggestion was nutrition labels\footnote{Participants based in the US were likely referring to the US Nutrition Facts labels: https://www.fda.gov/food/nutrition-education-resources-materials/new-nutrition-facts-label.}, as they are common on food packaging. P10 elaborates: ``When you buy anything from the grocery or the supermarket, if it's packaged, it usually comes with all that information and that information can be pretty accurate and handy.'' Those who wanted to extract more details for users also explored the idea of having the model identify specific ingredients, but acknowledged that it is a difficult task with photos. Other approaches consisted of combining 2 or more of the provided class combinations (P19, P24, P26), and highlighting substances that commonly trigger health problems such as sugar and cholesterol (P16). While these new classes may not all be realistic, it provides UXPs with a foundation for discussions with AI stakeholders, as P14 pointed out: ``I think it would be helpful for me to discuss with [the data scientist on my team] on what's the right categorization, and what are we trying to achieve.'' 

Through exploration, participants derived two common goals for the POC: accuracy and flexibility. Although no participants gave a concrete definition of accuracy, they generally referred to accuracy as the ability of the model to correctly classify images via the probability scores on an evaluated image (see Fig. \ref{f:tm-interface}). Many participants considered accuracy to be paramount for the app, so much so that P17 declared ``if it's not accurate, then the rest [of the app] is meaningless.'' P5 thought it was important to show promise of higher accuracy, even if current accuracy may not be ideal: ``hopefully I'll help them see that this thing will be pretty accurate and even if it's not accurate now, we'll figure out a way to make it more accurate in the future as we launch this product.'' P4 associated higher accuracy with higher satisfaction of user needs: ``my assumption would be that if it's more accurate, it's more likely to meet user needs [...] users are looking for something that's accurate.'' Regarding flexibility, participants wanted the app to account for a wide range of user preferences and customization, such as medical conditions, dietary restrictions, height, weight, and more. P21 saw the need for ``significant user customization'' in order for the app to function cross-culturally, even suggesting that users can define their own classes for the model. However, they also noted a tension where users may just want a model to perform reasonably ``out-of-the-box'' with minimal extra training. 

Taking the aforementioned goals into consideration, we asked participants if they consider their app to be a good use of AI in the first place. Most (18) stated yes, but interestingly, not many justified their answer with accuracy or flexibility. Instead, they primarily cited convenience and reduced need for manual data logging. 7 participants were ambivalent about the use of AI. Their reasons for positive views of AI were similar---AI accelerates the process of tracking and logging food. Their negative views stemmed from the lack of accuracy and flexibility, along with inability to manually correct misclassified results, subjectivity in classes such as \textsc{healthy} and \textsc{unhealthy}, and more. P10 and P23 also questioned whether AI was really necessary to achieve user goals:
\begin{quote}
    \textit{``Users can [manually sort] through very simple pre-created labels. Maybe it's even faster. So going back to the design process that the tool proposed, it is not really doing or helping the user to achieve their goals in any better way, so why are we even using those models?''} (P10)
\end{quote}
Finally, the remaining two participants did not consider the app to be an effective use of AI, as the AI was not accurate nor flexible enough to account for diverse use cases.

\subsubsection{Visualizing and Exploring Training Data Was Highly Valued}
\label{s:training-data}

Many participants considered it important to explore the training data. As P7 indicated: ``If [users] were to just see the input [images], then I think it's hard to imagine the output.'' P23 stated that ``it was less important to me to actually build a model and test it than it was to look at the data set that I was working with''. P23 went on to indicate that seeing the training data allowed them to anticipate the model's erroneous judgements. P15 said that inspecting the quality of training images can help them design guides for users to take better ones when using the app:

\begin{quote}
    \textit{``The training images that you had also have different scenarios where it's like a little bit darker or like it's really zoomed in. That was that was good for me to see because we can train our users on how to take their picture.''}
\end{quote}

Viewing training data prompted P17 to reflect on the assumptions behind the data: ``You have to trust the data source right, like [what if] there's inherent bias, or there's ignorance, or who's deciding [healthy and unhealthy], is it nutritionists?'' For P14, viewing the data acts as a probe for improving model performance: ``Currently in fruits, we have 14 sample images, but I'm curious if I added like 100 images, would [the model] be more accurate?'' 

Some participants 
actually considered viewing training data to be a sufficient proxy to the act of model training itself.
For example, P10 stated that upon inspecting the training sets: `` kind of know how it was labeled so I kind of expect what comes out. I can already visualize what will come out from this model, so [training the model] doesn't really do too much.'' however, later on, P10 acknowledged that training a model can still be useful in exposing model errors. 

We note that in our study, training data was visual and only contained 300 images, making it relatively easy to skim and explore, but this is not the case for all data formats. Given the importance of interacting with training data, we encourage researchers to consider what interfaces may be suitable for exploration of other data formats, such as text or audio, 

\subsubsection{Probability Scores Were Easily Misunderstood}
As seen in Fig. \ref{f:tm-interface} D, Teachable Machine provides a visualization of category probabilities for images evaluated by a trained model. Many participants found this visualization useful for better understanding how the model worked: 

\begin{quote}
    \textit{``You have all those bar graphs that show you how much percentage is classified as fruit or vegetable, or healthy and unhealthy. To see how it works in the backend was really useful because that gave me a clear mental model of how I would want to show it to the users''} (P26).
\end{quote}

That said, we were surprised to find that many participants misinterpreted these output probability scores---instead of reading the number as a percentage with which the model is confident in a particular class over others, participants read it as the percentage \textit{of the image} that the model identified as belonging to that class. P3 aptly summarized this confusion: 

\begin{quote}
    \textit{``I definitely was confused about whether it was the amount of the prediction of whether this will be accurate, or whether it was a split in a certain diet. And a lot of the times I went to the latter.''}
\end{quote}

This confusion may have affected participants' evaluation of model accuracy. For example, P5 saw narrowly distributed probabilities and believed the model was unable to detect elements within an image: 

\begin{quote}
    \textit{``There was an egg photo when I was doing the test, and it was a hundred percent protein. Sure the egg is protein, but that picture also had other components to it that this [model] wasn't able to tell, so from that aspect I think it's only partially useful.''}
\end{quote}

P23, who was aware of both interpretations of the visualization and was unclear about which one to accept, mentioned that their design would depend on the interpretation. If it is the output probability score, they would allow the user to review the scores and ensure the class with the highest score aligns with the user's expectations, and if it is a percentage of the image, they would incorporate a UI that shows the average percentage makeup of ingredients in a user's diet over time. Similarly, once P21 realized what the correct interpretation was, they wanted to focus more on communicating the model output to users so they will not be misled. We learn from this misunderstanding that conceptual misinterpretations can still exist in a simplified tool such as Teachable Machine (which all participants claimed to be intuitive and easy-to-understand); thus, correct interpretations should not be assumed.

\subsubsection{Explainability Benefits Are Carried Downstream}
\label{s:downstream-effects}
Teachable Machine, at the time of writing, does not offer explanations for model outputs beyond the visualization of class probabilities. Most participants wanted more explainability from Teachable Machine to better understand how to improve the model, gain more trust in the model, and design more informative user experiences for users of their app. P4 brought up feature relevance and believed that explainability is needed only when the model errs (which is almost always): ``if everything works as I imagine, I won't bother to look at what features are being valued; it's like when it's not being accurate that I want to understand why.'' P5 felt a bit helpless without concrete guidance on how to improve the model upon encountering errors: 

\begin{quote}
    \textit{``I wish there was some sort of suggestion about [whether] this [is] something wrong with the image or there are other actions I can take. Because I feel like there was no call to action, I'm just left to figure out this myself.''}
\end{quote}


When asked about why explainability was important to them, participants pointed out that the richer understanding enabled by explanations did not benefit only themselves, but allowed them to pass the benefits emerging from that enriched understanding to end-users. P15 summarizes this:

\begin{quote}
    \textit{``For me to actually understand how [the model] works would help me design the screens to be like, okay, this is information that we can show, but this is what we need to require from the user. Also understanding [model] limitations can also help [identify] any guardrails that we need to design for the user.''}
\end{quote}

P21, who wanted users to be able to correct model output and give feedback to the system, agreed that explainability can help them ``know more about how [the model] works in order to help users through the correction process.'' P19 was particularly interested in the language used to explain model decisions, and wanted Teachable Machine to provide examples of such strings so they can relay them to the user:

\begin{quote}
    \textit{``The strings are kind of written in a way so that it's like, `we estimate your meal based off like to be blah blah blah,' like there's that kind of language, where it shows that this was based off of a model. I think knowing more about the machine learning process can be better might help inform some of the strings that actually appear in the app.''}
\end{quote}

Overall, participants showed that the benefits they reap from explanations in their tools can make their way downstream and allow end-users of their designs to access those benefits as well. Not only that, explanations for UXPs can also serve as design inspiration for how AI concepts may be communicated to end-users. 

\subsubsection{Summary}
Our UXPs benefitted from hands-on experience with IML in many ways. They were able to quickly validate and/or reject assumptions and hypotheses they had about AI, learn about model errors and limitations, and anticipate potential biases. While previous work showed some similar benefits by allowing designers to explore a pre-trained model, our work engages UXPs in an end-to-end (albeit simplified) model training workflow and differs in two key areas. First, we found that the ability to interactively test different combinations of classes during training allowed UXPs to better reason about model alignment with user needs and identify new potential for the model. Second, we found that participants often relied on inspecting training data for model understanding, implying that high visibility of training data aids sensemaking. Additionally, we identified a common misinterpretation of the probability score visualizations and found that it can affect perceptions of model performance as well as interface design choices. Finally, we posit that explainability in IML tools not only benefits UXPs in the design process, but also end-users as UXPs forward those benefits downstream.

\subsection{Designing For User Interaction}


\subsubsection{Output Visualization From Teachable Machine Informed Design of Outputs in POC} 
\label{s:output-vis-informs}
Teachable Machine's visualization of class probabilities in the output module (see Fig. \ref{f:tm-interface} D) was well-liked by participants despite some misinterpretations. We observed from participants' POC proposals that most (22) took inspiration from the visualization for their designs, integrating UI elements with class probabilities (although the information may be in a different form, such as a pie chart). A subset of those participants (17) incorporated the bar chart visualization directly, either drawing their own replica of it or using a screenshot from Teachable Machine. Participants were also able to derive new UI elements from the visualization to address some of its perceived shortcomings. For example, P14 converted all class probability scores to checkmarks (if the percentage is non-zero) to avoid subjecting the user to excessive numerical details. P27 included in their app a ``health score,'' a composite metric that measures the balance between classes in the 5-class model. 

Although participants introduced new UIs and metrics, they were still based on the initial Teachable Machine visualization; we did not see evidence of any participants venturing beyond the initial visualization to explore alternative ones. Nor did they need to, as P16 indicated: ``[once we have the probability scores] we don't have to think of other ways of showing it to the user.'' This reveals that Teachable Machine's interface can in fact \textit{constrain the design space} of the POC. That is, by providing only one output visualization, Teachable Machine offers UXPs a low-hanging fruit to design off of, thus limiting exploration of other possible designs. 

\subsubsection{Participants Designed Guides To Help Users Maximize Utility of AI}
Experimenting with models and viewing training data allowed UX practitioners to derive design affordances to help users make better use of AI in the POC. Many recognized the quality of the image was a large factor in prediction accuracy, so they incorporated guides for users to take better photos. P14 suggested notifying the user of low-light conditions: ``tell the user [in the camera viewfinder] that the lighting is really bad, so improve your lighting, so we can make better predictions.'' P8 also recognized this, and took inspiration from the iPhone camera's night mode: ``whenever you're doing night mode on an iPhone, it's like `you need a little bit more light' or something; it's guiding you.'' On a higher level, P9 educated users on what acceptable images look like: ``I added a lot of educational components, just to tell them what pictures will generate better results.'' P3 also informed the user that the result is ``only in response to the information that you are giving us'' to set performance expectations. Besides lighting, P25 hypothesized that accuracy can be improved by guiding users to take photos of individual ingredients and designed their interface around that. 


In Section \ref{s:training-data}, we saw how visualizing and exploring training data enhanced UX practitioners' understanding of AI. Here, we see how practitioners pass that understanding along to users by scaffolding user interactions with informative guides.

\subsubsection{Participants Placed Great Emphasis on Manual Assignments and Corrections}
\label{s:manual-touch}
Most participants wanted their users (and themselves) to manually adjust model outputs to give feedback to the model. This is aligned with the ML subfield of \textit{active learning}, but active learning is not yet commonly deployed \cite{budd2021active}. Teachable Machine did not implement active learning. Primary reasons UX practitioners wanted this feature were to improve user trust in the model (which P18 considered to be ``the most challenging part [when working] with machine learning''), further customization for individual users, and address unanticipated errors. P19 summarized this desire as ``high touch opportunities'' when talking about the adjustment UI they included in their design:

\begin{quote}
    \textit{``At the end of the day, regardless of what kind of information the app presents to the user---so let's say, nutritional breakdown---users are able to still modify and have some high touch opportunities to [interact with] that kind of information.''}
\end{quote}

Models that accept feedback were also better aligned with participants' mental models of AI. Among others, P26 had a vision that corrections would result in a ``constantly improving model, or something that's just dynamically updating its results,'' echoing some expectations of AI mentioned in Section \ref{s:mental-models}. P2 envisioned a collaborative approach to improving model performance through manual correction: ``It makes sense if someone making corrections makes their data better, but in theory, it'd be also great if it would make someone else's [data] better.'' Besides users adjusting outputs, P21 believed it is important for UXPs to do the same, as it allows them to better envision how the process works ``in order to help users through the correction process.''

In addition to the adjustment of model outputs, some participants also saw the need for users to be able to create their own set of classes and train a personalized model around those classes, as P1 describes:

\begin{quote}
    \textit{``It would be interesting for users to be able to choose what categories they wanted to use. I know some people are really into micro and macro foods or whatever, and so, if you could choose that as the user can  take control of how they want to track their food, that could be really cool''}
\end{quote}

P21 argues that this style of customization is essential to for the app to function cross-culturally. They note that ``it sounds like from the persona, they really want [the model] to make good decisions off the shelf'' but they believe users need to ``start from scratch, with each user building their own dataset.'' P23 agrees: 

\begin{quote}
    \textit{``I can imagine, using a tool like [Teachable Machine] to permit a user to create their own categories, like I wouldn't I don't think I would want them to pre-train. I think I would just let a user take like a picture of all [their] meals that [they] eat for a week, upload those photos, and assign them a value that [they] care about.''}
\end{quote}

\subsubsection{Separation of AI and Non-AI User Experiences Can Be Blurry}
Some participants, such as P9, recognized that AI presents novel design challenges but found it difficult to isolate AI considerations from the rest of the design process. P15 mentioned that they would ideally work with a collaborator in charge of other (non-AI) parts of the app to better integrate the AI features into a broader user journey. P16 agreed that they ``couldn't randomly just take [AI explorations] out from that entire user journey and design with it.'' However, this was not the perspective of all participants. Many said that although Teachable Machine gave them confidence to reason about and communicate AI concepts, they still expect their ML team or data scientist collaborators to handle the AI separately. P8, P17, and P26 all mentioned that they do not see a difference between designing with AI compared to another technology, as both are merely the ``backend'' and are not of concern to user-facing ``frontend'' interfaces.

\begin{quote}
    \textit{``I don't think there was anything that would make it more challenging by it having an AI interface. I'm like looking at my [user] flow and I'm not seeing anything different. You launch it, you take a photo, you review whatever it is, and then you're done. The AI is in the back end. Nothing is different for the user''} (P17).
\end{quote}

P6, who has prior AI experience, also echoes this, but recognized that there should be deeper consideration of user consent and awareness of the presence of AI:

\begin{quote}
    \textit{``When pure UX is concerned, I don't see it as significantly different from any other user experience. I think that the user consent and awareness [of AI being used] needs to be a little bit more, but apart from that, in terms of experience, it needs to be like anything else I guess.''}
\end{quote}

This disagreement is particularly insightful because it highlights the desire and expectation to abstract key AI characteristics away from end-users. That is, despite AI being inherently probabilistic, participants still attempt to conform AI user experiences to those of traditional algorithms. A forced alignment of fundamentally different experiences can be the source of novel challenges (and confusion) for participants. For example, the aforementioned misinterpretation of probability scores may be due to the assumption that the AI model operates with full certainty. 
As AI-enabled probabilistic interfaces become more widespread, we see great importance in educating UXPs to explicitly disambiguate AI and non-AI experiences.

\subsubsection{Summary}
With IML, participants saw numerous design opportunities to enhance end-user interactions with AI. The visualizations shown to them offered inspiration for ways to display model output to users, but also appeared to limit further exploration of possibilities. To help users better leverage capabilities of the model, participants offered guides in their designs for users to take higher quality photos, allowed users to adjust model output to provide feedback to the model, and even gave users full customization over their model classes. That said, some still perceived AI considerations to be separate from UX ones, which we posit can induce friction in the design process.

\subsection{Ethical Considerations}
\subsubsection{Societal and Topical Risks}
Participants identified numerous ethical considerations with regard to societal impacts and potential harms of applying AI to the food domain. Many mentioned that the app may cause or perpetuate existing eating disorders and encourage unhealthy relationships with food. P7 was particularly concerned about how numerical metrics in the app can gamify eating: ``If it was gamified in the way that I proposed, then it could cause a few unhealthy habits in terms of obsession with achieving a certain score.'' P1 acknowledged that food is a socially sensitive topic and the app may not be well-suited for some: ``I feel like these apps can be a danger zone for some people who are maybe more prone to like anorexia or things like that [...] food can be a very sensitive topic for people with eating disorders, or a lot of people have shame around food.'' Similarly, P12 stated that model classes may fail to account for dietary restrictions and can force an undesirable balance: 

\begin{quote}
    \textit{``Maybe someone cannot eat meat or dairy or maybe someone cannot digest grain or something. Giving those classes [in the 5-class model] to people, it forces them to like focus on that balance, and maybe [that] could be dangerous for someone if they have certain diseases or conditions or risks.''}
\end{quote}

P26 saw potential in users' (or even data annotators') dietary biases or opinions being perpetuated to other users if the app allowed for such behaviour:

\begin{quote}
    \textit{``Maybe there's someone who's vegan and they think eating meat is not particularly healthy, and that is completely up to their own beliefs. They may classify the model to be unhealthy, and that could further inform the consumers of the app that if they're eating meat, then it's unhealthy or dangerous.''}
\end{quote}

Some also cited the lack of cultural flexibility as a concern, particularly with the 2-class model. P21 did not want the model to interpret healthiness from a culturally biased dataset: ``what about cultures that don't need a lot of dairy but eat a lot of rice? Like their stuff's going to be wrong if we go from just this dataset.'' P25 agreed with this concern: ``if they train the machine learning algorithm with mostly Western food, when the user takes photos of other like ethnic foods, maybe it will be detected as unhealthy.'' Likewise, many were concerned about the subjectivity of \textit{healthy} and \textit{unhealthy} in the 2-class model, once again referencing the lack of flexibility: 

\begin{quote}
    \textit{``Our [persona] was an athlete in training, but [the app] can just be really useful to pregnant people, or people recovering from being ill. They might have different needs and different amounts of those [food] groups. Healthy or unhealthy seemed a bit too broad-stroked.'' (P2)}
\end{quote}

P10 also surfaced the same subjectivity concerns in the 3-class model, claiming that \textsc{appetizers}, \textsc{entrees}, and \textsc{desserts} were arbitrary and culturally-dependent categorizations.

\subsubsection{Technical Risks}
Besides societal and domain-specific risks, participants also recognized risks stemming from AI's technical capabilities (or lack thereof). Inaccuracies in classifications were commonly mentioned as a risk. P1 and P8 both saw the potential of model misclassifications to promote over- or under-eating of certain foods. P8 and P9 were concerned that inaccuracies will sacrifice user trust and fail to deliver on the app's promises: ``if they're expecting to be able to quickly take a photo and for it to be right, then you know we really have to deliver on that and not lie to them'' (P8). Due to inevitable errors, P3 thought it was important to ``communicate that [the app] not going to be 100\% accurate, and always consult as somebody who is a [health] expert.'' P16 agrees that the app should not be decisive: ``ethically I think it's important to tell [users] that this is not the gospel truth, that this is just like a rough estimate of what could be a healthier way of eating.'' Indeed, leaning away from decisiveness is what makes in-app recommendations a double-edged sword for users, as P25 realized:

\begin{quote}
    \textit{``I was planning to [combine] those machine learning outputs with other resources to provide actionable and insightful suggestions to users [...] but machine learning trying to provide too many specific diet suggestions may trigger some people's anxiety about their eating habits.''}
\end{quote}

Participants were also aware of privacy risks that can come with AI. P2 and P11 both recognized that it was important for users to know where the images were stored and processed, and whether storage was cloud-based or local. P15 and P24 both designed permission notifications as part of their onboarding user flow. P2 added that permissions should 1) disclose legal boundaries and 2) ask about users' comfort in sharing various data. Additionally, P9 said it was important to disclose any data sharing that may happen with 3rd parties, such as selling datasets to companies that may give users diet-specific advertisements. 

\subsubsection{Summary}
Participants actively engaged in the consideration of ethical issues in their designs, both ones directly related to AI as well as ones that are a by-product of AI in the domain-specific context of food. Some thought these issues could be mitigated and AI would still be a fitting technology for the app. Others, such as P17 and P21, believed that risks overwhelmed benefits and that AI was a not viable solution.

\section{Discussion}
%
%
%
%
We used Teachable Machine and a prepared design task to better understand how UXPs might design and propose a proof-of-concept, ML-enabled interface. The combination enabled UXPs to more deeply reason about how AI can align with end-users needs, how they might design interfaces for effective interaction with an AI model, and prompted them to consider AI-related ethics and risks. 
Our discussion reflects on the promise and limitations of IML in the context of a UX design process. We then introduce \textit{research-informed machine teaching} (RIMT) as a conceptual guide for UXPs and discuss the potential of RIMT to mitigate IML's limitations in UX settings.

\subsection{IML Promotes Experience-Based Enrichment of Human-AI Guidelines}
\label{s:iml-guidelines}
Amershi et al. \cite{amershi2019guidelines} developed a set of 18 generally applicable design guidelines for human-AI interaction. Similar guidelines have also been conceived by research teams at Google \cite{pair-guidebook} and Apple \cite{apple-guidelines}. Throughout our study, we observed that many of these guidelines were naturally recognized by participants through a combination of our task-driven, IML-supported design exercise and their UX frame of mind. Participants then attempted to apply those guidelines, enriched by the surrounding context of the design problem, to the study design task. While a couple participants were aware of such guidelines, most started our study with no AI design experience and did not mention any prior knowledge of human-AI guidelines. We reviewed participants' ``lessons learned'' and design choices from the sessions (which we more generally call ``insights'') and mapped them onto Amershi et al.'s guidelines \cite{amershi2019guidelines}. We identified references to 12 of the original 18 guidelines. The mapping is shown in Table \ref{t:insights-guidelines}, where we only included insights from those with no prior AI design experience as examples.

\begin{table*}[h]
\centering
\caption{Alignment of Amershi et al. \cite{amershi2019guidelines}'s guidelines for human-AI interaction with insights UXPs shared with us after completing the design session. Note that all UXPs listed in this table had no prior experience designing with AI.}
    \begin{tabular}{p{2cm} p{5cm} p{7cm}}
    \toprule 
    ID (from \cite{amershi2019guidelines}) & Guideline & Example UXP Insight From Design Session\\
    \midrule 
    
    G1 & Make clear what the system can do. & Informing users that AI may make errors, particularly in earlier periods of usage (P18). \\
    G2 & Make clear how well the system can do what it can do. & Use hedging language and tell users to consult health experts for conclusive advice (P3). \\
    G4 & Show contextually relevant information. & Guide users to take better-lit photos in low-light environments (P8). \\
    G5 & Match relevant social norms. & Avoid subjective labelling of meal courses as they may vary across cultures (P19). \\
    G6 & Mitigate social biases. & Avoid irrelevant model classes for users with dietary restrictions---e.g. having a dairy class for vegan users  (P16). \\
    G8 & Support efficient dismissal. & Allow users to manually label images that the AI inaccurately classified (P4). \\
    G9 & Support efficient correction. & Provide sliders in the model output UI so users can adjust as needed (P15). \\
    G10 & Scope services when in doubt. & Reduce or remove recommendations on how users should eat and live (P13). \\
    G11 & Make clear why the system did what it did. & Incorporate short strings that briefly describe features the model is using to make a decision (P21).\\
    G13 & Learn from user behaviour. & Observe signs of dietary restrictions early on and ask users if they would like to eliminate unobserved classes (P12). \\
    G15 & Encourage granular feedback. & Provide an interface for users to adjust numerical model outputs (P2). \\
    G17 & Provide global controls. & Allow users to define, label, and train on their own classes (P23). \\
    \bottomrule
    \end{tabular}
    
    \label{t:insights-guidelines}
\end{table*}

We dub this alignment as \textit{experience-based enrichment} of human-AI guidelines. That is, rather than learning about the guidelines by simply reading them, UXPs understand and apply these guidelines through personal, empirical experience, against a backdrop of context-specific design problems and user needs. Providing UXPs with a list of human-AI guidelines for static consumption may result in a shallow understanding sufficient to resolve short-term challenges and questions. However, by offering UXPs experience-based approaches to realize those guidelines, we can enrich existing mental models with experiential realization (as proposed by constructivist approaches to learning \cite{piaget1976piaget, sarkar2016constructivist}) to deepen understanding of AI as a design material. This way, UXPs can be better prepared to apply and adapt their knowledge for more sophisticated, larger-scale challenges in the UX of AI.

\subsection{RIMT as a Conceptual Guide in UX}
As noted in Section \ref{s:manual-touch}, UXPs have a natural inclination towards active learning (a subarea of IML \cite{chung2020interactive}), because it more closely resembles the ``always-on'' human learning process. They envision users creating custom models based on their own data, categories, and classes. But this idea differs from active learning as defined by ML literature \cite{settles2012active, raghavan2006active, budd2021active} in a key way: users, not the algorithm, have control over label planning and selection. How can we better guide UXPs to design affordances for interactive model engagement while allowing users, rather than the model, to customize models?

We see promise in incorporating the paradigm of interactive machine teaching (IMT) into design tools\footnote{Here, we use ``design tools'' to broadly refer to tools UXPs may use at any stage of a design process, which may include tools like Teachable Machine. Our definition differentiates our implications from previous work, which considered a narrower idea of design tools, mostly focusing on those used for prototyping \cite{subramonyam2021protoai}.} to balance these desires. Many of our UXPs wanted more interactivity in their work with the AI/ML model, and at the same time, they wanted more information about the users, their needs, their goals, and their differing food or health contexts. This creates the opportunity to leverage UXPs' natural work practices of understanding users and apply those practices to the IMT concept of model teaching to guide explorations of ML as a design material. We call this conceptual guide \textit{research-informed machine teaching} (RIMT).

\begin{figure*}[h]
    \centering
    \includegraphics[width=1\textwidth]{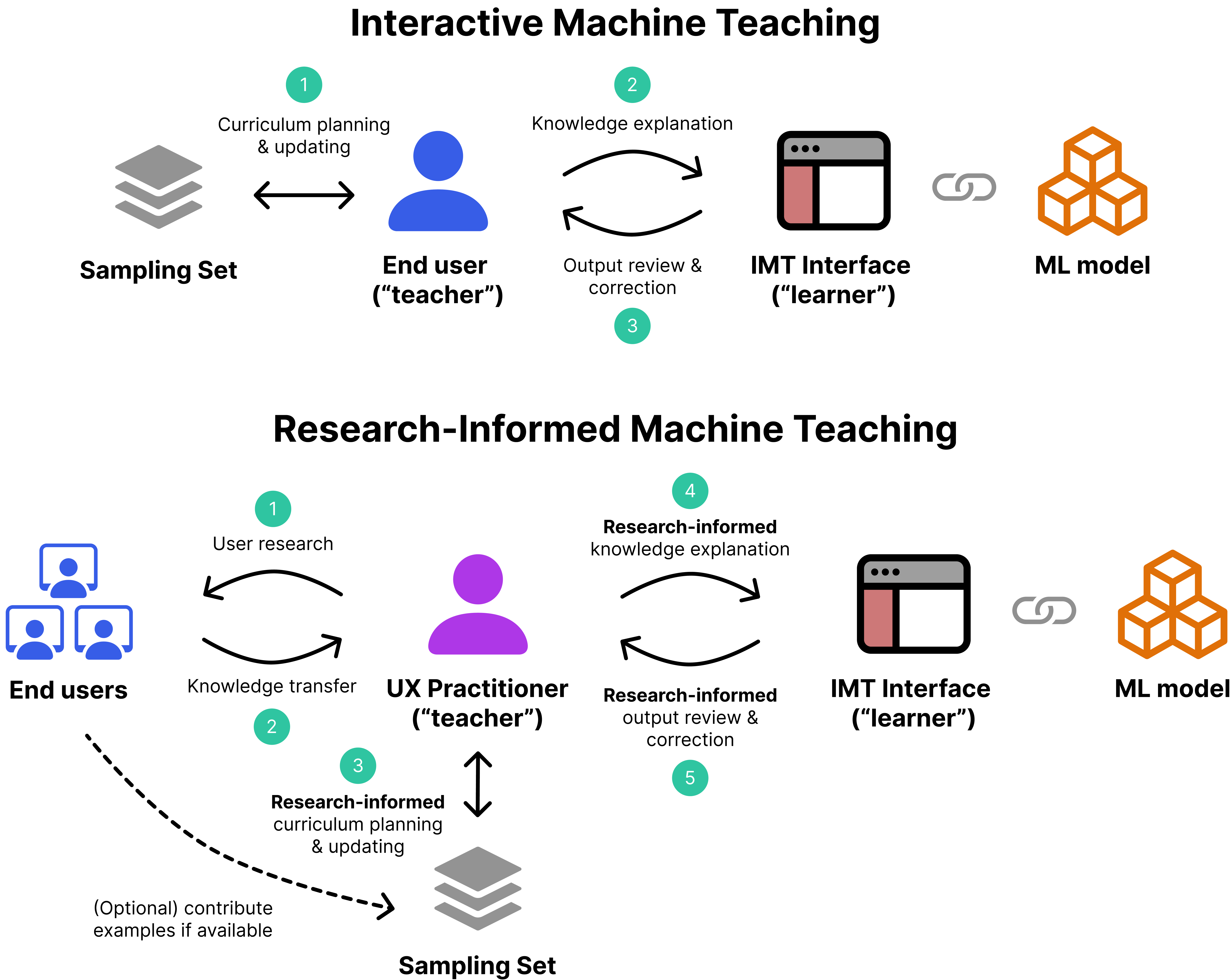}
    \caption{Comparison of IMT and RIMT. Note that both are iterative processes and the number markers denote order in which activities may be performed within one iteration.}
    \label{f:rimt}
    \Description{A comparison of the teaching loops of IMT and RIMT. The main difference lies in the inclusion of user research at the beginning of the process to obtain key information to be used in research informed teaching activities later on.}
\end{figure*}

RIMT is inspired by the IMT loop of planning, explaining, and reviewing (see Fig. \ref{f:rimt}). A key difference between RIMT and IMT is the employment of user research to inform activities in the loop. Rather than relying on self-contained knowledge to teach the model, UXPs acquire knowledge from their target end-user demographic through user research and use that knowledge as a proxy to teach the model on the end-user's behalf. For example, when teaching a binary classification model for \textsc{healthy} and \textsc{unhealthy} food, it is a prerequisite for the UXP to conduct sufficient user research to determine what exactly their end-users consider as \textsc{healthy} and \textsc{unhealthy}, before engaging in teaching activities. More fundamentally, the goals of RIMT and IMT are distinct. RIMT enables UXPs to bridge the gap between ML capabilities and end-user needs, while IMT is an interaction paradigm that allows end-users to create a model for their domain-specific tasks via a non-expert interface. RIMT does not necessarily imply that an IMT interface should be used in the eventual system---it is a design aid rather than an implementation technique. In fact, one may realize through user research that some end-user asks are unfit for IMT due to IMT's non-trivial teaching overhead---as P21 stated, end-users may want to use a model that just works ``out-of-the-box.'' The core value of RIMT lies in conceptually guiding UXPs to more deeply understand ML as a design material in tandem with user needs, and forming connections between the two. 

\subsection{Towards Harmonization of IML and RIMT in Design Tools}
We introduced RIMT as a conceptual guide, but note it awaits empirical validation in future work. Below, we offer one way to get started by operationalizing RIMT in design tools. We realize, however, that just like IMT, RIMT can demand considerably more time than the 30-second training times in Teachable Machine, as well as more attention from UXPs to actively label and correct examples. This may jeopardize the tight feedback loop essential to rapid prototyping. As such, we also consider possible interactions between RIMT and conventional IML. 


To combine RIMT's prioritization of human knowledge with the convenience of rapidly-trained, ``static'' models, we suggest the addition of \textit{teaching modes} in design tools. Teaching modes are environments that can be launched to situate UXPs in the teaching loop \cite{ramos2020imt} for both creation and refinement of models. They can launch a teaching mode from the beginning of their workflow if they wish to create a model from scratch, engaging in an iterative loop of research-informed planning, explaining, and reviewing in the typical RIMT fashion. They can also launch a teaching mode after the creation of a conventional model to refine and improve its performance. In this case, the curriculum (existing training task and data) has already been established by prior training, and the UXP reviews the existing curriculum to ensure they understand what it entails before adding new examples. From there, the UXP has entered the teaching loop and may engage in routine RIMT activities, with the exception of curriculum creation. They explain the new examples to the learning agent, review performance and correct when necessary, update the curriculum according to new research data, and repeat. To exit the teaching loop, the UXP deactivates the teaching mode and the model once again becomes unreceptive to feedback, permitting off-the-shelf use. It is essential for training modes to explicitly separate IML and RIMT as the two workflows employ fundamentally different approaches to knowledge extraction \cite{ramos2020imt}. 

We can draw an analogy between the two scenarios of training mode usage to teachers in a grade school system. When a UXP uses a teaching mode to train a model from scratch, they are like a full-time teacher who designs and leads an entire course, keeping close track of student progress throughout. When a user launches a teaching mode on a partially-trained model, they are like substitute a teacher who likely has less familiarity with the curriculum, but can still help advance students' knowledge. The ``full-time teacher'' here is not another human, but an automated ML pipeline. Of course, there is a key requirement in order for the latter scenario in our example to succeed: the substitute teacher has sufficient prerequisite knowledge and/or resources to become familiar with the curriculum on short notice. For more complex ML tasks where the automated teacher has constructed a curriculum beyond reasonable human understanding, launching a teaching mode on a partially trained model may be inadvisable. We see curriculum knowledge-sharing between multiple (human) teachers as a rich avenue for future work in (R)IMT.

We note that teaching modes may be actualized as human-AI guidelines instead of directly embedded into design tools. However, given our discussion of experience-based guideline enrichment and observed influences of design tools on designs for the end-user (see Section \ref{s:output-vis-informs}), we posit that teaching modes may have more impact when made accessible within design tools. 


\section{Limitations and Future Work}
Our research protocol was based on a simple, supervised image classification task in an experimental setting. Because of the visual nature of the data, participants were able to quickly glean preliminary insights and form hypotheses from training samples, which may not be the case with other forms of data such as text or audio. Future work may extend this study to more data formats and perhaps uncover additional desiderata around UXPs' exploration of non-visual data. Future work may also explore how UXPs handle training tasks that may require more specialized knowledge. The main focus of our design prompt was food, a universally understood and relatable topic, but many applications of ML in areas such as accessibility do not intrinsically contain an extensive body of shared experiences. In these cases, the RIMT curriculum may be difficult to grasp and it is therefore valuable to investigate how collaborating with domain experts can assist with curriculum development. Furthermore, as our study only involved supervised learning, we have yet to envision how IML- and RIMT-enabled design tools can look like for other ML techniques such as unsupervised and reinforcement learning.

Lastly, our analysis took place in an experimental setting in which UXPs were given a concrete task isolated from other (theoretical) team members. As some participants pointed out, collaboration with ML stakeholders is a necessary step in resolving ambiguities when the task is not so well-defined, as well as transitioning the app from a POC to real, usable technology. To complement current work in UXP collaboration with ML experts, future work may extend our analysis along a collaborative dimension and look at how IML- and RIMT-enabled design tools facilitate communication across UX and ML domain boundaries.

\section{Conclusion}
Advancements in ML motivate its increasing inclusion in user-facing applications. While prospects can be exciting, previous work has shown that UXPs face numerous challenges in working with ML as a design material, thus limiting the contribution of user-centered design perspectives in ML-enabled applications. We performed a contextual inquiry where we enabled UXPs to rapidly create and experiment with ML models as part of their UX workflows. We found that UXPs---even those with no prior ML exposure---were able to reason about ML and its interactions with end-users in sophisticated ways. We discuss the potential of RIMT in addressing some of IML's UX limitations, and how the two can co-exist in future design tools. As much of the current work in IML and IMT is centered around end-users, we are excited by their potential in UX to empower the creation and proliferation of human-centered ML interfaces.


\begin{acks}
We extend a warm thanks to all our participants and reviewers. We also thank Meena Muralikumar, Ruican Zhong, and Rock Pang for reading and providing feedback on drafts, as well as Gonzalo Ramos for engaging discussions about interactive machine teaching.
\end{acks}

\bibliographystyle{ACM-Reference-Format}
\bibliography{refs}




\end{document}